\documentclass[11pt,ams]{article}%
\usepackage{geometry}
\usepackage{dsfont}
\usepackage{amsmath}
\usepackage{amsfonts}
\usepackage{amssymb}
\usepackage{graphicx}%

\usepackage{float}

\begin{document}

\date{}
\title{\textbf{A study of the zero modes of the Faddeev-Popov operator in Euclidean Yang-Mills theories in the Landau gauge in $d=2,3,4$ dimensions }}
\author{\textbf{M.~A.~L.~Capri}$^{a}$\thanks{capri@ufrrj.br}\,\,,
\textbf{M.~S.~Guimaraes}$^{b}$\thanks{msguimaraes@uerj.br}\,\,,
\textbf{S.~P.~Sorella}$^{b}$\thanks{sorella@uerj.br}\ \thanks{Work supported by
FAPERJ, Funda{\c{c}}{\~{a}}o de Amparo {\`{a}} Pesquisa do Estado do Rio de
Janeiro, under the program \textit{Cientista do Nosso Estado}, E-26/101.578/2010.}\,\,,\,\,\textbf{D.~G.~Tedesco}$^{b}$\thanks{dgtedesco@uerj.br}\,\,\\[2mm]
\textit{{\small $^{a}$ UFRRJ $-$ Universidade Federal Rural do Rio de Janeiro}}\\
\textit{{\small Departamento de F\'{\i}sica $-$ Grupo de F\'{\i}sica Te\'{o}rica e Matem\'{a}tica F\'{\i}sica}}\\
\textit{{\small BR 465-07, 23890-971, Serop\'edica, RJ, Brasil.}}\\
\textit{{\small {$^{b}$ UERJ $-$ Universidade do Estado do Rio de
Janeiro}}}\\\textit{{\small {Instituto de F\'{\i}sica $-$
Departamento de F\'{\i}sica Te\'{o}rica}}}\\\textit{{\small {Rua
S{\~a}o Francisco Xavier 524, 20550-013 Maracan{\~a}, Rio de
Janeiro, RJ, Brasil.}}}$$}
\maketitle

\vspace{2cm}

\begin{abstract}
\noindent Examples of normalizable zero modes of the Faddeev-Popov operator in $SU(2)$ Euclidean Yang-Mills theories in the Landau gauge are constructed in $d=2,3,4$ dimensions.
\end{abstract}

\baselineskip=13pt

\newpage

\section{Introduction}

The understanding of the nonperturbative aspects responsible for the long distance behavior of confining Yang-Mills theories is one of the most challenging issues in theoretical physics. \\\\The common paradigm in the quantization of non-Abelian gauge theories relies on the Faddeev-Popov formula, which enables us to construct a meaningful version of the theory at the quantum level in the weak coupling regime, {\it i.e.} a theory which is renormalizable and unitary. We underline that unitarity is here referred to the case in which asymptotic fields can be safely introduced, so that a physical asymptotic subspace with positive norm can be constructed through the use of the BRST symmetry. The situation becomes very different when dealing with the infrared behavior of a confining theory as, for example, $QCD$, where asymptotic quark and gluon fields cannot be introduced due to color confinement. Despite the progress done in the last decades, a complete understanding of the behavior of non-Abelian gauge theories in the infrared regime accounting for color confinement is still lacking. \\\\One possible way to face this problem is to go back to the Faddeev-Popov quantization scheme and realize that, as it stands, it does not account for the existence of equivalent gauge field configurations fulfilling the same gauge fixing condition. This is the Gribov problem \cite{Gribov:1977wm}. Let us quote, as an example, the case of the Landau gauge, which has been intensively investigated in the last years. We remind that, according to Singer's result \cite{Singer:1978dk}, the existence of the Gribov phenomenon is a general feature of the gauge fixing procedure. Gribov copies have in fact been observed  in most of the employed covariant and renormalizable gauges, namely: the Landau gauge \cite{Gribov:1977wm, Henyey:1978qd, Sobreiro:2005ec,Maas:2005qt,Holdom:2009ws}, the class of general linear covariant gauges \cite{Sobreiro:2005vn},  the maximal Abelian gauges \cite{Bruckmann:2000xd,Guimaraes:2011sf}. Also, the Coulomb gauge provides an example of a non-covariant gauge exhibiting the Gribov phenomenon \cite{Ilderton:2007qy, Holdom:2010sg,Canfora:2011xd}.    The fact that the Faddeev-Popov formula is plagued by the existence of Gribov copies means that the gauge fixing procedure does not pick up a unique representative for each gauge orbit. As a consequence, the counting of the relevant   field configurations is not properly done. Although a complete resolution of the Gribov issue is not yet available, it has become now clear that taking into account the Gribov copies requires a modification of the Faddeev-Popov quantization formula in the infrared non-perturbative region, as  expressed by the so called Gribov-Zwanziger action \cite{Zwanziger:1988jt,Zwanziger:1989mf},  which enables us to restrict the domain of integration in the functional integral to the region in field space bounded by the first Gribov horizon. This restriction deeply modifies the behavior of the Green's functions of the theory in the infrared. As an example, the two-point gluon correlation function turns out to be suppressed in the infrared, exhibiting complex poles. As such, it cannot describe physical excitations, {\it i.e.} gluons are destabilized by the presence of the Gribov horizon and are not part of the physical spectrum, as expected in a confining Yang-Mills theory.  Recently, a refined version of the Gribov-Zwanziger theory has been established \cite{Dudal:2007cw,Dudal:2008sp,Dudal:2011gd}, through the introduction of dimension two condensates. We point out that the predictions of the refined version of the Gribov-Zwanziger theory turn out to be in remarkable agreement with lattice numerical simulations, both at the level of the gluon and ghost propagators, see \cite{Dudal:2010tf,Cucchieri:2011ig} and references therein, as well as at the level of the spectrum of the lightest glueball states \cite{Dudal:2010cd}. All these nontrivial results can be taken as evidence that the Gribov phenomenon plays a rather relevant role in the infrared physics of Yang-Mills theories. \\\\The aim of this work is that of pursuing the study of the issue of the Gribov copies, already started in \cite{Guimaraes:2011sf}. More specifically, we shall provide examples of infinite classes of renormalizable zero modes of the Faddeev-Popov operator in the Landau gauge, $\partial_i A^a_i=0$, $i=1,2,..,d$, in $d=2,3,4$ Euclidean dimensions. We shall  look thus at normalizable solutions of the equation
\begin{equation}
{\cal M}^{ab} \omega^b = 0  \;, \label{zm}
\end{equation}
where ${\cal M}^{ab}$ stands for the Faddeev-Popov operator in $d$ dimensions
\begin{equation}
{\cal M}^{ab} = - \left( \partial^2 \delta^{ab} + \varepsilon^{acb} A^c_i \partial_i \right)   \;, \label{fp}
\end{equation}
and $\varepsilon^{acb}$ are the structure constants of $SU(2)$.  Nontrivial solutions of the equation \eqref{zm} immediately lead to the existence of the Gribov copies. Considering in fact the two equivalent gauge configurations $({\hat A}^{a}_i, A^a_i)$
\begin{equation}
{\hat A}^{a}_i = A^a_i - \left( \delta^{ab} \partial_i + \varepsilon^{acb} A^c_i \right) \omega^b  \;, \label{gbc}
\end{equation}
it turns out that, due to eq.\eqref{zm},  both fields ${\hat A}^{a}_i$ and  $A^a_i$ fulfill the Landau gauge condition, {\it i.e.}
\begin{equation}
\partial_i {\hat A}^{a}_i = \partial_i A^a_i +  {\cal M}^{ab} \omega^b   = \partial_i A^a_i = 0  \;. \label{equiv}
\end{equation}
Therefore, the gauge configuration ${\hat A}^{a}_i$ is a Gribov copy of ${A}^{a}_i$. \\\\The strategy which we shall follow to obtain nontrivial normalizable solutions of  eq.\eqref{zm} is that outlined by Henyey in \cite{Henyey:1978qd}, and already successfully employed in  \cite{Guimaraes:2011sf}. Henyey's strategy amounts to find a parametrization of both zero mode $\omega^a$ and related field configuration $A^a_i$ in such a way that eq.\eqref{zm} becomes an algebraic equation enabling one to express the gauge field in terms of the function parametrizing the zero mode itself. Despite its simplicity, Henyey's framework turns out to be very powerful in order to provide examples of gauge copies with the desired asymptotic behavior for the resulting gauge field configuration. \\\\In what follows we shall consider two kinds of boundary conditions for the behavior of the gauge field at infinity. The first type is expressed by the requirement that the Yang-Mills action is finite, {\it i.e.}
\begin{equation}
S_{YM} = \frac{1}{4g^2} \int d^d x \; F^a_{ij} F^{a}_{ij}  < \infty \;. \label{fact}
\end{equation}
The finite action condition looks a natural requirement to be imposed, given the important fact that the action has the meaning of Boltzmann weight in the Euclidean functional integral. \\\\The second type of boundary condition which we shall discuss amounts to require that the behavior of the gauge field at infinity is such that the  Hilbert norm $||A||^2$ is finite \cite{Jackiw:1977ng}, namely
\begin{equation}
||A||^2 = \int d^dx\; A^a_i A^a_i  < \infty \;. \label{normmf}
\end{equation}
Condition \eqref{normmf} is stronger than the requirement of finite action, eq.\eqref{fact}. Field configurations fulfilling the condition of finite norm $||A||^2$ automatically lead to a finite action, the converse being not true, as it will be shown by explicit examples.  The relevance of the Hilbert norm $||A||^2$ relies on the fact that, in the Landau gauge, several properties of both Gribov region and fundamental modular region can be established by demanding that the functional
\begin{eqnarray}
f_A[u] & = & \int d^dx \; \left( A^u_i \right)^a \left( A^u_i \right)^a \;, \nonumber \\
A^u_i & = & u^{\dagger} A_i u + u^{\dagger} \partial_i u   \;, \label{fu}
\end{eqnarray}
is well defined and differentiable along  the gauge orbit of $A_i$ \cite{semenov,Dell'Antonio:1991xt,vanBaal:1991zw}. It turns out in fact that one possible way to introduce the Gribov region $\Omega$\footnote{We remind that the Gribov region $\Omega$ is defined as the set of field configurations fulfilling the Landau gauge condition and for which the Faddeev-Popov operator ${\cal M}^{ab}$  is strictly positive, namely
\begin{equation}
\Omega = \{ \;A^a_i\;;\;\; \partial_i A^a_i=0 \;, \;  {\cal M}^{ab} = - \left( \partial^2 \delta^{ab} + \varepsilon^{acb} A^c_i \partial_i \right) >0 \;\; \}  \;. \label{Om}
\end{equation} }
is to identify it with the set of all relative minima of the functional $f_A[u]$  in the space of the gauge orbits, while the fundamental modular region corresponds to the set of all absolute minima. The fundamental modular region is contained within the Gribov region and enjoys the property of being free from Gribov copies \cite{semenov,Dell'Antonio:1991xt,vanBaal:1991zw}. Though, a practical way of implementing the restriction to the fundamental modular region in the domain of integration in the functional integral is, at present,  not known. \\\\The present work is organized as follows. Sections 2,3,4 are devoted to construct explicit examples of infinite classes of normalizable zero modes of the Faddeev-Popov operator, whose corresponding gauge field configurations display an asymptotic behavior compatible with the requirements \eqref{fact},  \eqref{normmf} in $d=2,3,4$ dimensions, respectively. In Sect.5 a few concluding remarks are collected.

\section{Henyey's construction of zero modes of the Faddeev-Popov operator in the Landau Gauge in $d=2$ dimensions}

Let us begin by discussing the construction of an infinite class of normalizable zero modes of the Faddeev-Popov operator in $d=2$ dimensions. Let $A_i$, $i=1,2$, be a $SU(2)$ Lie-algebra valued gauge field in Euclidean space,
\begin{eqnarray}
A_i = A^a_i \left( \frac{\sigma^a}{2} \right) \;, \qquad [\sigma^a,\sigma^b] = 2 i\epsilon^{abc} \sigma^c \;, \label{gf}
\end{eqnarray}
where $\sigma^a$, $a=1,2,3$, denote the Pauli matrices. \\\\Constructing normalizable zero modes of the Faddeev-Popov operator amounts to find solutions of the equation
\begin{eqnarray}
\partial^2 \omega^a + \epsilon^{acb}A^{c}_{i} \partial_i \omega^b = 0 \;, \label{operator}
\end{eqnarray}
with
\begin{equation}
\int d^2x\; \omega^a \omega^a < \infty \;. \label{fom}
\end{equation}
Adopting polar coordinates ($r$,$\varphi$)\footnote{see (Appendix \ref{ap})}, the Landau gauge condition, $\partial_i A^a_i=0$,  can be fulfilled by setting
\begin{eqnarray}
 A_i^b = \delta^{b3}a_i \equiv \delta^{b3}\epsilon_{ij}x_j f(r) \;,  \label{b3}
\end{eqnarray}
so that eq.\eqref{operator}  becomes
\begin{eqnarray}
 \partial^2\omega^a +f(r) \epsilon^{3ac}\epsilon_{ij}x_i\partial_j\omega^c = 0 \;.  \label{bec}
\end{eqnarray}
The quantity $f(r)$ in eq.\eqref{b3} stands for an arbitrary function depending only on the radial coordinate $r$. Looking at the second term, one recognizes the angular momentum $L$
 \begin{eqnarray}
	 L= \epsilon_{ij}x_i\partial_j = \frac{\partial}{\partial \varphi} \;. \label{angm2}
\end{eqnarray}
Then, choosing the components of $\omega^a(r,\varphi)$ along the directions orthogonal to $A^a_i$ in color space, namely $\omega^a = (\omega^{(1)},\omega^{(2)},0)$, we obtain
\begin{eqnarray}
 \partial^2\omega^{(1)} +f(r) L\omega^{(2)} &=& 0\; \nonumber \\
 \partial^2\omega^{(2)} -f(r) L\omega^{(1)} &=& 0 \;. \label{2dsph}
\end{eqnarray}
We parametrize $(\omega^{(1)},\omega^{(2)})$ in the following way
\begin{eqnarray}
\omega^{(1)}(r,\varphi)& = & \sigma(r)\sin\varphi \;, \nonumber \\
\omega^{(2)}(r,\varphi) & = & \sigma(r)\cos\varphi \;. \label{param}
\end{eqnarray}
where $\sigma(r)$ is left free. As a consequence, eqs.\eqref{param} merge into a single equation, which reads
\begin{eqnarray}
 \sigma''(r) + \frac{1}{r}\sigma'(r) - \frac{1}{r^2}\sigma(r) = f(r)\sigma(r) \;. \label{singeq}
\end{eqnarray}
Expression \eqref{singeq} illustrates in a simple way Henyey's construction \cite{Henyey:1978qd}. Both gauge field, eq.\eqref{b3}, and zero mode, eq.\eqref{param}, are parametrized in such a way that eq.\eqref{singeq} allows us to express the function $f(r)$ in terms of the quantity $\sigma(r)$. More precisely,
making the change of variables
\begin{equation}
\sigma(r) = r \psi(r) \;, \label{ch}
\end{equation}
it follows that
\begin{eqnarray}
 f(r) = \frac{1}{\psi(r)}\left( \psi''(r) + \frac{3}{r}\psi'(r) \right) \;. \label{f2}
\end{eqnarray}
What remains to be done is to find a suitable function $\psi(r)$  which provides a normalizable zero mode $\omega^a$ and a corresponding gauge field $A_i^a$ with the desired asymptotic behavior. To that purpose we shall employ the following trial expression for $\psi(r)$:
\begin{eqnarray}
 \psi(r) = \frac{1}{r^m + \alpha} \;,  \label{psi}
\end{eqnarray}
where the exponent $m$ is positive and $\alpha$ is an arbitrary positive constant. From expression \eqref{f2}, for the function $f(r)$ one obtains
\begin{eqnarray}
 f(r) = \frac{1}{(r^m + \alpha)^2} \left( r^{2m-2}(m^2 - 2m) -\alpha r^{m-2}(m^2 + 2m)\right) \;. \label{expf}
\end{eqnarray}
Also, the norm of the zero mode $\omega^a$ is easily evaluated and turns out to be given by
\begin{eqnarray}
 \|\omega\| ^2 = \int d^2x\; \omega^a \omega^a \propto \int_0^\infty dr \;r \; \frac{r^2}{(r^m+\alpha)^2} \;. \label{norm}
\end{eqnarray}
This expression will be convergent for $m>2$. Note that in this case we have, asymptotically, $f(r)\sim 1/r^2$, and thus
\begin{equation}
A_i  \sim 1/r  \;, {\it \; \qquad for} \;\;\; r\rightarrow \infty  \;. \label{asympt}
\end{equation}
It is easy to check that the asymptotic behavior \eqref{asympt} leads to a finite Yang-Mills action. In fact, one observes that, since only the 3-color component of the field $A^a_i$ is nonvanishing, eq.\eqref{b3}, the field strength, $ F^{a}_{ij} = \partial_i A_j^a -\partial_j A_i^a + \epsilon^{abc}A_i^bA_j^c $, has a unique nontrivial component, namely
\begin{eqnarray}
 F^{3}_{ij} = \partial_i A_j^3 -\partial_j A_i^3 \;. \label{f3}
\end{eqnarray}
Therefore, making use of eq.\eqref{b3}, we have
\begin{eqnarray}
 F^{3}_{ij} = 2 \epsilon_{ij} f(r) + \epsilon_{ik} x^k x^j \frac{f'(r)}{r} - \epsilon_{jk} x^k x^i \frac{f'(r)}{r} \;. \label{f3exp}
\end{eqnarray}
Moreover, using the identity
\begin{eqnarray}
 \epsilon_{ik} x^j + \epsilon_{kj} x^i +\epsilon_{ji} x^k = 0 \;, \label{id}
\end{eqnarray}
we find
\begin{eqnarray}
 F^{3}_{ij} = \epsilon_{ij}( 2f(r) + rf'(r) ) \;. \label{finalf3}
\end{eqnarray}
Finally, for the Yang-Mills action one gets
\begin{eqnarray}
 S_{YM} = \frac{1}{4g^2}\int d^2x F^{3}_{ij}F^{3}_{ij} = \frac{\pi}{g^2}\int_0^\infty dr\, r ( 2f(r) + rf'(r) )^2 < \infty \;, \label{fYM}
\end{eqnarray}
which is obviously finite, as $f(r)\sim 1/r^2$. Observe, however, that this is not the case of the Hilbert norm $||A||^2$, which is logarithmic divergent, due to $A_i  \sim 1/r$. We see thus that expression  \eqref{psi} gives rise to an infinite class of normalizable zero modes of the Faddeev-Popov operator, parametrized by the exponent $m>2$, whose corresponding gauge field configurations yield a finite Yang-Mills action.

\section{Zero modes in $d=3$}

In order to construct examples of zero modes in $d=3$, we shall proceed by generalizing the set up outlined in the case $d=2$. Adopting polar coordinates ($r$,$\theta$,$\varphi$), as starting gauge configuration $A^a_i, i=1,2,3$, fulfilling the Landau condition, $\partial_i A^a_i=0$, we shall employ
\begin{eqnarray}
 A_i^c = \epsilon_{cij}x_j h(r) \;,\label{d3gaugeconf}
\end{eqnarray}
where the function $h(r)$ depends only on the radial coordinate $r$. The differential equation for the zero modes reads
\begin{eqnarray}
 \partial^2\omega^a -
h(r) \epsilon^{acb}\epsilon_{cij}x_i\partial_j\omega^b = 0 \;. \label{ddff}
\end{eqnarray}
As already observed in the previous section, one easily recognizes in the second term of eq.\eqref{ddff} the angular momentum $L^a$
 \begin{eqnarray}
 L^a= \epsilon^{aij}x_i\partial_j \;. \label{angm}
\end{eqnarray}
Moreover, similarly to the $d=2$ case, we choose $\omega^a(r,\varphi,\theta) = (\omega^{(1)},\omega^{(2)},0)$, so that eq.\eqref{ddff} gives
\begin{eqnarray}
 \partial^2\omega^{(1)} - h(r) L^3 \omega^{(2)} &=& 0 \;, \nonumber \\
 \partial^2\omega^{(2)} + h(r) L^3 \omega^{(1)} &=& 0 \;, \label{3dsph}
\end{eqnarray}
where
\begin{equation}
L^3 = \frac{\partial}{\partial \varphi}  \;. \label{l3}
\end{equation}
Therefore, setting
\begin{eqnarray}
 \omega^{(1)} (r,\varphi,\theta)&=\sigma(r)\sin\varphi \sin\theta \;,\nonumber\\
 \omega^{(2)} (r,\varphi,\theta)&=\sigma(r)\cos\varphi \sin\theta \;, \label{pmom}
\end{eqnarray}
it follows that eqs.(\ref{3dsph}) reduce to a single equation, given by
\begin{eqnarray}
 \sigma''(r) + \frac{2}{r}\sigma'(r) - \frac{2}{r^2}\sigma(r) = h(r)\sigma(r) \;. \label{eqh}
\end{eqnarray}
Adopting thus the change of  variables $\sigma(r) = r \psi(r)$, we obtain
\begin{eqnarray}
 h(r) = \frac{1}{\psi(r)}\left( \psi''(r) + \frac{4}{r}\psi'(r) \right) \;. \label{hh1}
\end{eqnarray}
In order to find an infinite class of normalizable zero mode we proceed as before and we make use of the function
\begin{eqnarray}
 \psi(r) = \frac{1}{r^p + \alpha} \;, \label{ppsi3}
\end{eqnarray}
with $p$ an arbitrary positive exponent and $\alpha$ a positive  constant. From eq.\eqref{hh1} we get
\begin{eqnarray}
 h(r) = \frac{1}{(r^p + \alpha)^2} \left( r^{2p-2}(p^2 - 3p) -\alpha r^{p-2}(p^2 + 3p)\right) \;. \label{hm}
\end{eqnarray}
The norm of the zero mode $\omega^a$ is easily computed and yields
\begin{eqnarray}
 \|\omega\| ^2 \propto \int_0^\infty dr \frac{r^4}{(r^p+\alpha)^2}  \;. \label{nrmh}
\end{eqnarray}
The integral of equation \eqref{nrmh} is convergent for $p>5/2$. Therefore, expressions  \eqref{pmom}  and  \eqref{ppsi3}  provide  an infinite class of normalizable zero modes, parametrized by the values of the exponent  $p>5/2$. \\\\Also, for a generic value of $p>5/2$, the asymptotic behavior of $h(r)$ in expression \eqref{hm} is $h(r)\sim 1/r^2$  for $r\rightarrow \infty$. As a consequence, the behavior of the corresponding gauge field $A^c_i$, eq.\eqref{d3gaugeconf}, is
\begin{equation}
A_i \sim \frac{1}{r}\;,  \qquad { for} \qquad  r\rightarrow \infty \;. \label{b3d}
\end{equation}
Again, similarly to the previous case in $d=2$, this gauge configuration \eqref{hm} gives rise to a finite Yang-Mills action. In fact for the field
strength $F^a_{ij}$ we get
\begin{eqnarray}
 F^{a}_{ij} = 2\epsilon_{aji} h(r) - \epsilon_{aik} x^k x^j \frac{h'(r)}{r} + \epsilon_{ajk} x^k x^i \frac{h'(r)}{r} +
\epsilon_{abc} \epsilon_{bik} \epsilon_{cjl} x^k x^l h(r)^2 \;. \label{fs3}
\end{eqnarray}
Making use of the algebraic identity
\begin{equation}
\epsilon_{ijk} x^m -  \epsilon_{jkm} x^i  + \epsilon_{kmi} x^j - \epsilon_{mij} x^k = 0  \;, \label{3deps}
\end{equation}
expression \eqref{fs3} becomes
\begin{eqnarray}
 F^{a}_{ij} = \epsilon_{aji} (2h(r) + rh'(r)) + \epsilon_{ijk} \left(h(r)^2 + \frac{h'(r)}{r}\right)x^ax^k.
\end{eqnarray}
Therefore, for the Yang-Mills action one finds
\begin{eqnarray}
\int d^3 x F^a_{ij}F^a_{ij} &= 4\pi \int dr r^2 \left[6\left( 2h(r) + rh'(r)\right)^2 - 4r^2\left( 2h(r) + rh'(r)\right) \left(\frac{h'(r)}{r} +h(r)^2 \right) \right.\nonumber\\
&\left. +  2r^4\left(\frac{h'(r)}{r} +h(r)^2 \right)^2\right],
\end{eqnarray}
which is easily seen to be finite for $p>5/2$. \\\\However, for a generic value of $p>5/2$, the asymptotic behavior \eqref{b3d} does not guarantee that the Hilbert norm  $||A||^2= \int d^3x A^a_i A^a_i $ is finite too. Nevertheless, for the particular value of $p=3$, the expression for $h(r)$ reduces to
\begin{equation}
h(r) \big|_{p=3} = -\frac{18 \alpha r}{(r^3+\alpha)^2}  \;, \label{hm3}
\end{equation}
so that
\begin{equation}
A_i \big|_{p=3} \sim \frac{1}{r^4}\;,  \qquad { for} \qquad r\rightarrow \infty \;,  \label{bm=3}
\end{equation}
yielding a finite Hilbert norm
\begin{equation}
||A||^2\big|_{p=3}= \int d^3x \left(A^a_i A^a_i \right) \big|_{p=3}  < \infty \;.   \label{f3aa}
\end{equation}
Evidently, the field configuration \eqref{hm3} produces a finite Yang-Mills action. Although the expression
\begin{equation}
 \psi(r) = \frac{1}{r^3 + \alpha} \;, \label{psi3}
\end{equation}
gives only a particular example of a normalizable zero mode whose corresponding gauge field, eq.\eqref{bm=3}, displays a finite norm $||A||^2$, it is relatively easy to construct now an infinite class of zero modes giving rise to a finite norm. To that purpose, we introduce the generalized function
\begin{eqnarray}
 {\tilde \psi}(r) = \frac{1}{r^3 + \beta r^n + \alpha} \;, \label{tildepsi}
\end{eqnarray}
where $\beta$ is a positive constant and $n$ a positive exponent such that $n<3$. From eq.\eqref{hh1}, for the function $h(r)$ we obtain now
\begin{eqnarray}
 {h(r)} &= \frac{1}{(r^3 + \beta r^n + \alpha)^2} \left[ \beta^2 r^{2n -2}(n^2 - 3n) + \beta r^{n+1}(9n-n^2 - 18)\right.\nonumber\\
 &\left. - \beta\alpha r^{n-2}(n^2+3n) - 18r\alpha\right].
\end{eqnarray}
We see that we must constrain $n\geq2$ in order to avoid a singularity at the origin. Asymptotically we have $h(r)\sim 1/r^{5-n}$ and thus $A_i  \sim 1/r^{4-n}$. As a consequence, the Hilbert norm $\|A\|^2$ will be finite for
$n<\frac{5}{2}$. We have thus succeeded in constructing an infinite class of zero modes, parametrized by the exponent  $2<n<\frac{5}{2}$ of eq.\eqref{tildepsi}, whose corresponding gauge fields have finite Hilbert norm $\|A\|^2$.

\section{The case $d=4$}
As last example, let us now face the case $d=4$. As before, we adopt spherical coordinates ($r$,$\alpha$,$\theta$,$\varphi$). As starting configuration fulfilling the Landau gauge, we take
\begin{eqnarray}
 A_\mu^c= \epsilon_{c\mu\nu4}x_\nu g(r)\;,  \qquad \partial_\mu A_\mu^c = 0\;, \qquad \mu=1,2,3,4 \;, \label{4dgf}
 \end{eqnarray}
 where $g(r)$ depends only on the radial coordinate and  $\epsilon_{\mu\nu\rho\sigma}$ stands for the Levi-Civita antisymmetric tensor. Therefore, choosing the zero mode  $\omega^a (r,\alpha,\theta,\varphi)$ in the usual way, namely
\begin{eqnarray}
 \omega^a (r,\alpha,\theta,\varphi)= (\omega^{(1)},\omega^{(2)}, 0) \;, \label{zmd4}
\end{eqnarray}
it follows that $(\omega^{(1)},\omega^{(2)})$ obey the differential equations
\begin{eqnarray}
 \partial^2 \omega^{(1)} + g(r) \epsilon_{3\mu\nu4} x_\mu\partial_\nu\omega^{(2)} = 0 \;, \nonumber\\
 \partial^2 \omega^{(2)} - g(r) \epsilon_{3\mu\nu4} x_\mu\partial_\nu\omega^{(1)} = 0 \;. \label{4d_fp_op}
\end{eqnarray}
Once more, one recognizes the appearance of the angular momentum in eqs.\eqref{4d_fp_op}, {\it i.e.}
\begin{eqnarray}
 \epsilon_{3\mu\nu4} x_\mu\partial_\nu = x\partial_y - y\partial_x = \frac{\partial}{\partial\varphi} \;.
\end{eqnarray}
Eqs.(\ref{4d_fp_op}) reduce thus to
\begin{eqnarray}
 \partial^2 \omega^{(1)} + g(r) \frac{\partial}{\partial\varphi} \omega^{(2)} = 0 \;, \nonumber \\
 \partial^2 \omega^{(2)}  - g(r) \frac{\partial}{\partial\varphi}\omega^{(1)} = 0 \;. \label{4dsph}
\end{eqnarray}
Generalizing the same procedure adopted in the previous cases $d=2,3$, we parametrize the zero mode $\omega^a$ as:
\begin{eqnarray}
\omega^a (r,\alpha,\theta,\varphi)= (\sigma(r)\sin\theta\sin\alpha\sin\varphi,\sigma(r)\sin\theta\sin\alpha\cos\varphi, 0) \;. \label{pd4}
\end{eqnarray}
Equations (\ref{4dsph}) become a single equation, given by
\begin{eqnarray}
 \sigma''(r) + \frac{3}{r}\sigma'(r) - \frac{3}{r^2}\sigma(r) = g(r)\sigma(r) \;. \label{seq4d}
\end{eqnarray}
Employing the change of variables $\sigma(r) = r \psi(r)$, eq.\eqref{seq4d}  yields
\begin{eqnarray}
 g(r) = \frac{1}{\psi(r)}\left( \psi''(r) + \frac{5}{r}\psi'(r) \right) \;. \label{g4d}
\end{eqnarray}
As usual, we set
\begin{eqnarray}
 \psi(r) = \frac{1}{r^q + \alpha} \;, \label{c}
\end{eqnarray}
where $q$ is a positive exponent and $\alpha$ a positive constant. From \eqref{g4d}, for the function $g(r)$ we can write
\begin{eqnarray}
 g(r) = \frac{1}{(r^q + \alpha)^2} \left( r^{2q-2}(q^2 - 4q) -\alpha r^{q-2}(q^2 + 4q)\right) \;. \label{gexpr4}
\end{eqnarray}
For the norm of the zero mode,  we have
\begin{eqnarray}
 \|\omega\| ^2 \propto \int_0^\infty dr \frac{r^5}{(r^q+\alpha)^2}  \;, \label{intd4}
\end{eqnarray}
which is finite for $q>3$.  Expression \eqref{c} provides thus an infinite class of normalizable zero modes. \\\\Concerning now the function $g(r)$ in  eq.\eqref{gexpr4}, it behaves as $g(r) \sim 1/r^2$ for $r \rightarrow \infty$, giving rise to a gauge field $A\sim 1/r$. These configurations do not have finite norm $\|A\|^2$, although yield a finite action. In fact,
in this case, for the field strength one has
\begin{eqnarray}
 F^{a}_{ij} = 2\epsilon_{aji4} g(r) - \epsilon_{aik4} \;x^k x^j \frac{g'(r)}{r} + \epsilon_{ajk4}\; x^k x^i \frac{g'(r)}{r} +
\epsilon_{abc}\; \epsilon_{bik4}\; \epsilon_{cjl4} x^k x^l g(r)^2 \;, \label{fst}
\end{eqnarray}
which, upon contraction of the indices, becomes
\begin{eqnarray}
 F^{a}_{ij} = \epsilon_{aji4}\; (2g(r) + rg'(r)) + \epsilon_{ijk4}\; (g(r)^2 + \frac{g'(r)}{r})x^ax^k \;. \label{fst4}
\end{eqnarray}
Therefore, for the Yang-Mills action we obtain
\begin{eqnarray}
\int d^4 x F^a_{ij}F^a_{ij} &= 2\pi^2 \int dr r^3 \left[6\left( 2g(r) + rg'(r)\right)^2 - 4r^2\left( 2g(r) + rg'(r)\right) \left(\frac{g'(r)}{r} +g(r)^2 \right) \right.\nonumber\\
&\left. +  2r^4\left(\frac{g'(r)}{r} +g(r)^2 \right)\right] \;,  \label{ym4d}
\end{eqnarray}
which is easily seen to be finite, as $g(r) \sim 1/r^2$. \\\\Let us end this section by showing that, as in the case $d=3$, also in $d=4$ it turns out to be possible to find an infinite class of zero modes exhibiting finite Hilbert norm  $\|A\|^2$.  To that purpose, we first remark that, for the particular case $q = 4$,  expression  \eqref{gexpr4} becomes
\begin{equation}
g(r) \big|_{q=4} = -\frac{32 \alpha r^2}{(r^4+\alpha)^2}  \;, \label{hm4}
\end{equation}
so that $g\sim1/r^6$ for $r \rightarrow \infty$. As such, the corresponding gauge field behaves as $A_i \sim 1/r^5$, exhibiting thus finite norm $\|A\|^2$. As done in the previous case $d=3$, we consider now the generalized expression
\begin{eqnarray}
{\hat  \psi}(r) = \frac{1}{r^4 + \beta r^n + \alpha} \;, \label{gen4}
\end{eqnarray}
where $n$ is a positive exponent smaller than 4, $n < 4$, and $\beta, \alpha$ are positive constants. From eq.\eqref{g4d}, the corresponding expression for $g(r)$ is found to be
\begin{eqnarray}
 g(r) = \frac{1}{(r^4 + \beta r^n + \alpha)^2} &\{ \beta^2 r^{2n-2}(2n^2-5n) + \beta r^{n+2}(12n - n^2 -32) \nonumber \\
&+ r^{n-2}\beta\alpha(\beta n -5n - \beta n^2) - 32\alpha r^2 \} \;. \label{gm=4}
\end{eqnarray}
We see that, as in the case $d=3$, we must constrain $n\geq2$ in order to avoid the appearance of a singularity at the origin.  Asymptotically,  we have $g(r)\sim 1/r^{6-n}$ and thus $A_i  \sim 1/r^{5-n}$. As a consequence, the Hilbert norm $\|A\|^2$ will be finite  for $n<3$. Thus, also in $d=4$, we have been able to construct  an infinite class of zero modes, parametrized by the exponent  $2\leq n<3$ of eqs.\eqref{gen4},\eqref{gm=4},  whose corresponding gauge fields have finite Hilbert norm $\|A\|^2$.

\section{Conclusion}

In this work we have pursued the investigation of the issue of the Gribov copies. In particular, we have been able to provide examples of infinite classes of normalizable zero modes of the Faddeev-Popov operator in the Landau gauge, in $d=2,3,4$ dimensions, and for the gauge group $SU(2)$. \\\\Two boundary conditions for the behavior of the gauge field at infinity have been considered, namely: the requirement of finite Yang-Mills action,  {\it i.e.} $ S_{YM} = \frac{1}{4g^2}\int d^4 x F^a_{ij}F^a_{ij} < \infty $, and that of finite Hilbert norm,   {\it i.e.} $\|A\|^2< \infty$. The requirement of finite Hilbert norm   is stronger than that of finite Yang-Mills action. Gauge configurations leading to a finite norm automatically give rise to a finite action, the converse being not true. Nevertheless, in both cases, we have been able to provide in a rather simple way explicit examples of infinite classes of normalizable zero modes with the desired behavior for the gauge field at infinity. Here,  Henyey's set up \cite{Henyey:1978qd} has been proven to be particularly useful.  \\\\Let us conclude with a few remarks.

\begin{itemize}

\item the examples of the zero modes and related gauge field configurations which we have been able to construct in a rather easy way give a simple confirmation of the relevance and, at the same time, of the complexity of the issue of the Gribov copies for a correct quantization of Yang-Mills theories.

\item it is worth to point out that the Gribov region $\Omega$
\begin{equation}
\Omega = \{ \;A^a_i\;;\;\; \partial_i A^a_i=0 \;, \;  {\cal M}^{ab} = - \left( \partial^2 \delta^{ab} + \varepsilon^{acb} A^c_i \partial_i \right) >0 \;\; \}  \;, \label{Omf}
\end{equation}
can be introduced without any specific reference to a particular behavior of the gauge field at infinity. It is apparent in fact that the region $\Omega$ can be introduced in a way which is perfectly compatible with both boundary conditions which we have taken into account. In this sense, the restriction to that region, as implemented by the Gribov-Zwanziger action \cite{Gribov:1977wm,Zwanziger:1988jt,Zwanziger:1989mf}, has to be regarded as an important step towards a correct quantization of Yang-Mills theories. Certainly, the restriction to the region $\Omega$ enables one to eliminate all Gribov copies related to zero modes of the Faddeev-Popov operator.

\item As we have already underlined, several properties of the Gribov region $\Omega$, eq.\eqref{Omf}, have been established \cite{semenov,Dell'Antonio:1991xt,vanBaal:1991zw}, namely
\begin{itemize}
\item {\it i)} the region $\Omega$ is convex and bounded in all directions in field space,

\item {\it ii)} every gauge orbit crosses $\Omega$ at least once.
\end{itemize}
In particular, property {\it ii)} provides a support to the original Gribov proposal of restricting the domain of integration in the functional integral to the region $\Omega$. This restriction is achieved by the introduction of the Gribov-Zwanziger action \cite{Gribov:1977wm,Zwanziger:1988jt,Zwanziger:1989mf}, which can be summarized by the following formula
\begin{equation}
\int_{\Omega} [DA]\; \delta({\partial A}) \;det({\cal M}) \; e^{-S_{YM}} = \int [DA]\; \delta({\partial A}) \;det({\cal M}) \; e^{-\left( S_{YM}+S_h-\int d^4x 4\gamma^4(N^2-1) \right)  } \;, \label{fi}
\end{equation}
where $S_h$ is called horizon function and is given by the nonlocal expression
\begin{eqnarray}
S_h & = & \gamma^4 \int d^4x\; h(x) \;, \nonumber  \\
h(x) & = & g^2 \int d^4y f^{bal} A^a_\mu(x) ({\cal M}^{-1})^{lm}(x,y) f^{bkm} A^k_{\mu}(y) \;. \label{hz}
\end{eqnarray}
The massive parameter $\gamma$ is the Gribov parameter, defined in a self-consistent way through the gap equation  \cite{Zwanziger:1988jt,Zwanziger:1989mf}
\begin{equation}
\langle h(x) \rangle = 4 \gamma^4 (N^2-1) \;. \label{gpeq}
\end{equation}
The parameter $\gamma$ encodes the presence of the Gribov horizon, {\it i.e.} it originates from the restriction of the domain of integration in the functional integral to the region $\Omega$. As such, $\gamma$ can be seen as arising from the elimination in the functional measure of the field configurations which lie outside of the region $\Omega$, and which are necessarily Gribov copies of configurations belonging to $\Omega$.  As shown in  \cite{Zwanziger:1988jt,Zwanziger:1989mf}, the nonlocal horizon function implementing the restriction to $\Omega$ can be cast in a local form through the introduction of  a suitable set of localizing fields. Remarkably, the resulting local action turns out to be renormalizable to all orders. Recently,
a refined version of the Gribov-Zwanziger action has been worked out \cite{Dudal:2007cw,Dudal:2008sp,Dudal:2011gd}. The predictions obtained from the Refined Gribov-Zwanziger action  turns out to be in very good agreement with the lattice data on the gluon and ghost propagators, see  \cite{Dudal:2010tf,Cucchieri:2011ig} and references therein, as well as on the spectrum of the lightest glueballs \cite{Dudal:2010cd}. These preliminary results can be taken as evidence of the important role played by the Gribov issue towards a satisfactory understanding of the infrared behavior of confining Yang-Mills theories.

\item Finally, we should touch on the fact that the restriction to the Gribov region $\Omega$ is not yet sufficient to eliminate all possible copies. Additional copies are still present inside the region $\Omega$. As already mentioned, there exists a smaller region, contained inside the Gribov region, which is free from Gribov copies. This region is known as the Fundamental Modular Region \cite{semenov,Dell'Antonio:1991xt,vanBaal:1991zw}.  It is known that part of the boundary of the Gribov region is common to  the Fundamental Modular Region. In principle, the restriction of the domain of integration to the Fundamental Modular Region should provide a quantization formula for Yang-Mills theories free from any Gribov ambiguities.  Unfortunately, unlike the case of the Gribov region $\Omega$, a local and renormalizable action implementing the restriction to the Fundamental Modular Region is not available.  In view of these remarks, it is safe to state that the possible physical relevance of the field configurations which we have investigated in this paper could be asserted only after a detailed study of their relationship with the Fundamental Modular Region, a rather difficult task which is out of the aim of the present  work which attempts at providing examples of the zero modes of the Faddeev-Popov operator.  For a better idea of the relevance of this issue we remind the reader to the lattice investigations given in \cite{Greensite:2004ur,Sternbeck:2005vs}.

\end{itemize}

\section*{Acknowledgments}
The Conselho Nacional de Desenvolvimento Cient\'{\i}fico e
Tecnol\'{o}gico (CNPq-Brazil), the Faperj, Funda{\c{c}}{\~{a}}o de
Amparo {\`{a}} Pesquisa do Estado do Rio de Janeiro, the Latin
American Center for Physics (CLAF), the SR2-UERJ,  the
Coordena{\c{c}}{\~{a}}o de Aperfei{\c{c}}oamento de Pessoal de
N{\'{\i}}vel Superior (CAPES)  are gratefully acknowledged.

\section*{Appendix}\label{ap}

For the benefit of the reader, we remind here some basic expressions in polar coordinates in $d=2,3,4$ dimensions.
\subsection*{d=2}

\begin{itemize}
 \item polar coordinates

\begin{eqnarray}
x&=&r \cos\varphi \nonumber \\
y&=&r \sin\varphi
\end{eqnarray}

where $\varphi$ ranges over $[0,2\pi]$

\item orthonormal basis of unit vectors

\begin{eqnarray}
\hat{e}_r &=& \hat{i} \cos\varphi  + \hat{j}\sin\varphi  \nonumber \\
\hat{e}_\varphi &=& -\hat{i}\sin\varphi + \hat{j}\cos\varphi
\end{eqnarray}

\item differential operators

gradient
\begin{equation}
\vec{\nabla}f = \hat{e}_r \frac{\partial f}{\partial r} + \hat{e}_\varphi\frac{1}{r}\frac{\partial f}{\partial \varphi}
\end{equation}

Laplacian

\begin{equation}
\partial^2 f \equiv {\nabla}^2 f = \frac{1}{r} \frac{\partial}{\partial r} \left( r \frac{\partial f}{\partial r}\right)+\frac{1}{r^2}\frac{\partial^2 f}{\partial \varphi^2}
\end{equation}

\end{itemize}

\subsection*{d=3}
\begin{itemize}

\item the spherical coordinates are  given by
\begin{eqnarray}
x&=& r \cos\varphi \sin\theta  \nonumber \\
y&=& r \sin\varphi \sin\theta  \nonumber \\
z&=& r \cos\theta  \nonumber \\
\end{eqnarray}

where $\varphi$ ranges over $[0,2\pi]$ and $\theta$ ranges over $[0,\pi]$

\item basis of orthonormal vectors

\begin{eqnarray}
\hat{e}_r &=& \hat{i}\sin\theta\cos\varphi  + \hat{j}\sin\theta\sin\varphi  + \hat{k}\cos\theta \nonumber \\
\hat{e}_\theta &=& \hat{i}\cos\theta\cos\varphi + \hat{j}\cos\theta\sin\varphi - \hat{k}\sin\theta \nonumber \\
\hat{e}_\varphi &=& -\hat{i}  \sin\varphi + \hat{j}  \cos\varphi
\end{eqnarray}

\item differential operators

gradient
\begin{equation}
\vec{\nabla}f = \hat{e}_r \frac{\partial f}{\partial r} + \hat{e}_\theta \frac{1}{r}\frac{\partial f}{\partial \theta} + \hat{e}_\varphi \frac{1}{r\sin\theta}\frac{\partial f}{\partial \varphi}
\end{equation}

Laplacian

\begin{equation}
\partial^2 f \equiv \nabla^2 f = \frac{1}{r^2} \frac{\partial}{\partial r}\left(r^2\frac{\partial f}{\partial r}\right) + \frac{1}{r^2 \sin\theta} \frac{\partial}{\partial \theta}\left(\sin\theta\frac{\partial f}{\partial \theta}\right) + \frac{1}{r^2 \sin^2\theta}   \frac{\partial^2f}{\partial \varphi^2}
\end{equation}

\end{itemize}

\subsection*{d=4}

\begin{itemize}

\item in four dimensions, for the spherical coordinates we have
\begin{eqnarray}
x&=& r \cos\varphi \sin\theta \sin\alpha \nonumber \\
y&=& r \sin\varphi \sin\theta \sin\alpha \nonumber \\
z&=& r \cos\theta \sin\alpha \nonumber \\
t&=& r \cos\alpha
\end{eqnarray}

 where $\varphi$ ranges over $[0,2\pi]$, while ($\theta$,$\alpha$) range over $[0,\pi]$

\item basis of orthonormal vectors

\begin{eqnarray}
\hat{e}_r &=& \hat{i}\sin\theta\cos\varphi\sin\alpha + \hat{j} \sin\theta\sin\varphi\sin\alpha + \hat{k} \cos\theta\sin\alpha + \hat{t}\cos\alpha \nonumber \\
\hat{e}_\theta &=& \hat{i}\cos\theta\cos\varphi + \hat{j}\cos\theta\sin\varphi - \hat{k} \sin\theta  \nonumber \\
\hat{e}_\varphi &=& -\hat{i}\sin\varphi + \hat{j}\cos\varphi \nonumber \\
\hat{e}_\alpha &=& \hat{i}\sin\theta\cos\varphi\cos\alpha + \hat{j}\sin\theta\sin\varphi\cos\alpha + \hat{k} \cos\theta\cos\alpha- \hat{t}\sin\alpha \;\;\;\;
\end{eqnarray}

\item differential operators

gradient
\begin{equation}
\vec{\nabla}f = \hat{e}_r \frac{\partial f}{\partial r} + \hat{e}_\alpha \frac{1}{r}\frac{\partial f}{\partial \alpha} + \hat{e}_\theta \frac{1}{r\sin\alpha}\frac{\partial f}{\partial \theta} + \hat{e}_\varphi \frac{1}{r\sin\alpha\sin\theta}\frac{\partial f}{\partial \varphi}
\end{equation}

Laplacian

\begin{equation}
\partial^2 f\equiv  \nabla^2 f = \frac{1}{r^3} \frac{\partial}{\partial r}\left(r^3\frac{\partial f}{\partial r}\right) + \frac{1}{r^2 \sin^2\alpha} \frac{\partial}{\partial \alpha}\left(\sin^2\alpha\frac{\partial f}{\partial \alpha}\right) + \frac{1}{r^2 \sin^2\alpha} \left(  \frac{1}{\sin\theta}\frac{\partial}{\partial \theta}\left(\sin\theta\frac{\partial f}{\partial \theta}\right)+ \frac{1}{\sin^2\theta}\frac{\partial^2f}{\partial\varphi^2}\right)
\end{equation}

\end{itemize}

\end{document}